\documentclass[%
reprint,showpacs,preprintnumbers,
amsmath,amssymb,
aps,
prd,
]{revtex4-1}

\usepackage{graphicx}

\begin{document}

\preprint{Imperial/TP/10/DJW/02}
\title{Soliton form factors from lattice simulations}
\author{Arttu Rajantie, David J. Weir}
\affiliation{%
Theoretical Physics Group, Blackett Laboratory, Imperial College London, London SW7 2AZ, U.K.
}%

\date{\today}% It is always \today, today,
             %  but any date may be explicitly specified

\begin{abstract}
The form factor provides a convenient way to describe properties of topological solitons in the full quantum 
theory, when semiclassical concepts are not applicable. It is demonstrated that the form factor can be 
calculated numerically using lattice Monte Carlo simulations. The approach is very general and 
can be applied to essentially any type of soliton. The technique is illustrated by calculating  the kink form factor near the critical point in 1+1-dimensional scalar field theory. As expected from universality arguments, the result agrees with the exactly calculable scaling form factor of the two-dimensional Ising model.
\end{abstract}

\maketitle

%\tableofcontents

Topological solitons play an important role in a wide range of physical systems~\cite{Rajaraman1987,vshel,Manton:2004tk}, and they have been studied extensively both experimentally and theoretically. Although in many of these systems quantum mechanical effects are significant, theoretical studies have been mainly limited to the classical limit in all but the simplest cases. In principle, one can use perturbation theory to calculate `semiclassical' quantum corrections to classical quantities~\cite{Dashen:1974cj}. However, this only works when the quantum effects are small and in practice can only be used for very simple models~\cite{Kiselev:1988gf}. A fully quantum mechanical analysis requires not only a new calculational method but also a different set of observables; the simplest classical observables do not have well-defined quantum mechanical counterparts.

In this paper we investigate one such observable, the soliton form factor. The form factor is a fully non-perturbative quantum observable which does not rely on any semiclassical concepts, and it can be defined for any soliton in an analogous way. Form factors are used in many areas of physics to characterize properties of quantum objects, from atomic~\cite{Born} and nuclear physics~\cite{Jelley} to integrable systems~\cite{Smirnov:1992vz}. The soliton form factor describes the scattering of a particle with a soliton; it can be loosely interpreted as (the Fourier transform of) the soliton profile in the quantum theory. It is therefore the most natural quantum observable beyond the soliton mass, and it carries a large amount of non-trivial information about the soliton and its interactions. By studying the form factor, one can therefore move away completely from semiclassical ideas of soliton shape to work with fully non-perturbative results for excitations and interactions.

Choosing a concrete example with both nontrivial critical behaviour and well-understood semiclassical limits, we shall focus on the kink form factor~\cite{Goldstone:1974gf,Jackiw:1975im,Mussardo:2003ji,Mussardo:2005dx,Mussardo:2006iv} in 1+1-dimensional field theory. Previously~\cite{Rajantie:2009bk}, we studied semiclassical aspects of kinks with lattice Monte Carlo simulations by measuring the field correlation function in the presence of a kink. Here we show how that the same approach can be further developed to calculate the kink form factor in a fully non-perturbative way. We demonstrate this by carrying out simulations near the critical point, where we find excellent agreement with exact results from the two-dimensional Ising model, as predicted by universality arguments. This approach can be generalized to solitons in other theories.

Let us consider a theory with a real scalar field $\phi(t,x)$ and kinks in 1+1 dimensions.
Our discussion will be valid for any such theory, but
as a concrete example, we use the $\lambda\phi^4$ model with Lagrangian
\begin{equation}
\label{eq:lagrangian}
 \mathcal{L} = \frac{1}{2}(\partial_\mu \phi)(\partial^\mu \phi) + \frac{m^2}{2} \phi^2 - \frac{\lambda}{4!} \phi^4.
\end{equation}
For fixed coupling $\lambda$ there is a critical mass parameter value $m_c^2$ below which the 
${\mathbb Z}_2$ symmetry of this theory is spontaneously broken, and the scalar field has
a vacuum expectation value $\langle\phi\rangle=\pm v$.
In the classical theory $m_c^2=0$ and $v=m\sqrt{6/\lambda}$.
The kink corresponds to a state which interpolates between one vacuum on one side and the other vacuum on the other side.

We assume that the system is in a state with one kink.
Classically, this simply corresponds to the exact kink solution
$
\phi_{\rm kink}(x) = v\tanh(m x/\sqrt{2}).
$
In the quantum theory, the same can be achieved by imposing twisted antiperiodic boundary conditions in the spatial direction~\cite{Groeneveld:1980tt}. In fact, this only restricts the number of kinks to odd values, but states with more than one kink are exponentially suppressed in the infinite volume limit.  Note that this way of preparing the system preserves translation invariance and is fully non-perturbative as it makes no reference to any classical background configuration.

The ground state $|0\rangle$  of this one-kink sector of the theory corresponds to a kink in a momentum eigenstate with zero momentum, and its energy $E_0$ is therefore simply the kink mass $M$.
This state is therefore translation invariant. 
Above this, the spectrum consists of moving kink states $|k\rangle$ with momentum $k$ and energy $E_k=\sqrt{k^2+M^2}$,
up to the energy $E_{\rm exc}$  of the first excited state of the kink.
Above $M+m$, there are also states with one or more free scalar particles. 

We want to calculate the kink form factor $f(k,k')$, which is defined as the matrix element
\begin{equation}
\label{equ:fdef}
 f(k,k')
=\frac{1}{v} \left< k' \right| \hat\phi(0) \left| k \right>,
\end{equation}
where we have scaled it by the vacuum expectation value $v$ to make it independent of the field normalisation,
and the momentum states $|k\rangle$ have the Lorentz invariant
normalisation 
\begin{equation}
\langle k'|k\rangle=2\pi\delta(k-k')E_k.
\end{equation}
Lorentz invariance of the theory implies that, when expressed in terms of rapidities $\beta_k=\mathrm{arcsinh} \; k/M$,
the form factor is a function of the rapidity difference only~\cite{Jackiw:1975im}, $f(k,k')=f(\beta_k-\beta_{k'})$.

Semiclassically, the kink form factor is given by the Fourier transform of the static kink solution~\cite{Goldstone:1974gf,Mussardo:2003ji},
\begin{equation}
\label{eq:ftkink}
f_{\rm cl}(\beta)=\frac{4}{3}i\pi v^2\frac{1}{\sinh\frac{2}{3}\pi v^2\beta}.
\end{equation}
This means that even in the quantum theory the form factor can be thought of as the effective kink profile. However, this interpretation should not be taken literally because, as always, there are many quantum observables that have the same semiclassical limit.
The semiclassical approximation is valid at weak coupling. In our model (\ref{eq:lagrangian}), the dimensionless coupling is $\lambda/m^2=6/v^2$, thus weak coupling implies a large vacuum expectation value.

What makes the theory (\ref{eq:lagrangian}) a particularly useful test bed is that the form factor is also known exactly at strong coupling, by which we mean near the critical point $m^2\approx m_c^2$ in the quantum theory. 
The theory is in the same universality class as the two-dimensional Ising model, and near the critical point the
form factor should approach the exact Ising model result~\cite{Berg:1978sw,Yurov:1990kv},
\begin{equation}
\label{equ:fIsing}
f_{\rm Ising}(\beta) =  i  \coth \frac{\beta}{2}.
\end{equation}

Matrix elements like (\ref{equ:fdef}) cannot be computed directly using Monte Carlo simulations. Instead, 
the basic observable is the field correlation function, which we consider in the ground state $|0\rangle$ of the one-kink sector.
We calculate it in momentum space, taking the Fourier transform in space but not in time,
and write a spectral expansion in terms of energy eigenstates $|\alpha\rangle$ with energies $E_\alpha$,
\begin{equation}
\langle\phi(0,k)\phi(t,q)\rangle=\sum_\alpha\frac{\langle 0|\hat\phi(k)|\alpha\rangle
\langle\alpha|\hat\phi(q)|0\rangle}{\langle0|0\rangle}
e^{it(E_\alpha-E_0)}.
\end{equation}

Lattice Monte Carlo simulations are carried out in Euclidean space, which is obtained by carrying out a Wick rotation
$t\rightarrow it$. This does not affect the coefficients of the spectral expansion, but the exponentials become real,
\begin{equation}
\label{equ:spectral}
\langle\phi(0,k)\phi(t,q)\rangle=\sum_\alpha
\frac{\langle 0|\hat\phi(k)|\alpha\rangle\langle\alpha|\hat\phi(q)|0\rangle}{\langle0|0\rangle}
e^{-t(E_\alpha-E_0)}.
\end{equation}
At long enough time separation,
\begin{equation}
\label{eq:timesepcondition}
t\gg \frac{1}{E_{\rm exc}-E_0},
\end{equation}
the dominant contribution comes from the single-particle moving kink states $|k\rangle$. For them, the coefficient of the expansion is essentially the form factor, because
\begin{equation}
\langle k'|\hat\phi(q)|k\rangle=vf(k,k')2\pi\delta(k-q-k').
\end{equation}
The momentum conservation delta function restricts the expansion to only states with overall momentum $k$,
and therefore we have
\begin{eqnarray}
\langle\phi(0,k)\phi(t,q)\rangle&=&\frac{2\pi\delta(k+q)}{L}\frac{v^2|f(k,0)|^2}{E_kE_0}
e^{-t(\sqrt{k^2+M^2}-M)}
\nonumber\\&&+O\left(e^{-t(E_{\rm exc}-M)}\right),
\end{eqnarray}
where $L$ is the spatial length of the system, and we have used $\langle 0|0\rangle=L E_0$ as implied by our normalisation.

Furthermore, the Euclidean spacetime is necessarily finite in actual Monte Carlo simulations. We 
assume periodic boundary conditions in the time direction, and denote the length of the system by $T$.
In the periodic Euclidean time, the field correlator is
\begin{equation}
\label{eq:bigtrace}
\left< \phi(0,k)\phi(t,q) \right> = \frac{\text{Tr}\hat{U}(T-t) \hat\phi(q) \hat{U}(t) \hat\phi(k)}{\text{Tr} \hat{U}(T)},
\end{equation}
where $\hat{U}(t)=\exp(-\hat{H}t)$ is the Euclidean time evolution operator.
As in Eq.~(\ref{equ:spectral}), at long enough time separations the only contribution comes from single-particle kink momentum 
eigenstates $|k\rangle$, so we can approximate the trace in Eq.~(\ref{eq:bigtrace}) by an integral over them,
\begin{equation}
\label{equ:fullres}
\left< \phi(0,k)\phi(t,q) \right> = \frac
{\int \frac{dk'}{2\pi E_{k'}}\langle k'|\hat{U}(T-t) \hat\phi(q) \hat{U}(t) \hat\phi(k)|k'\rangle}
{\int\frac{dk'}{2\pi E_{k'}}\langle k'| \hat{U}(T)|k'\rangle}.
\end{equation}
Using
\begin{equation}
\langle k'|\hat{U}(t)|k\rangle=2\pi\delta(k-k')E_ke^{-E_kt},
\end{equation}
we can write the denominator as
\begin{equation}
\label{equ:denominator}
\int\frac{dk'}{2\pi E_{k'}}\langle k'| \hat{U}(T)|k'\rangle=L\int\frac{dk'}{2\pi}e^{-E_{k'}T},
\end{equation}
Inserting complete sets of momentum eigenstates, the numerator becomes
\begin{multline}
\label{equ:numerator}
\int \frac{dk'}{2\pi E_{k'}}\langle k'|\hat{U}(T-t) \hat\phi(q) \hat{U}(t) \hat\phi(k)|k'\rangle
\\
=2\pi\delta(q+k)\int  \frac{dk'}{2\pi}
\frac{v^2|f(k'-k,k')|^2}{E_{k'-k}E_{k'}}
e^{-E_{k'}(T-t)-E_{k'-k}t}.
\end{multline}

\begin{figure}[t]
\includegraphics[scale=0.66666]{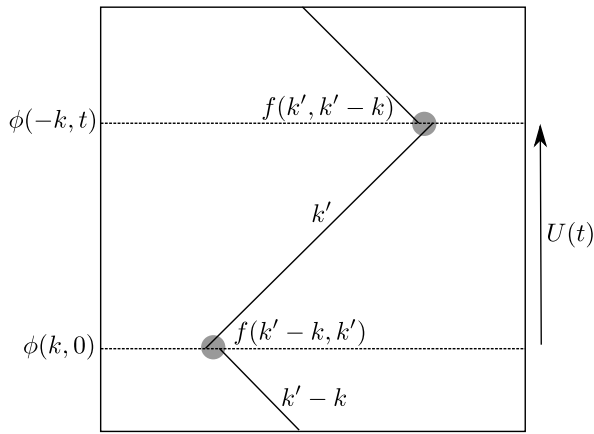}
\caption{\label{fig:cartoon} Illustration corresponding to Eq.~(\ref{equ:numerator}). The kink is constrained by periodic boundary conditions in the Euclidean time direction, so the worldline must match up at either end of the lattice. The defect is not point-like, and has a form factor which is represented here by the interaction between the scalar and the defect having a finite size.}
\end{figure}

As illustrated in Fig.~\ref{fig:cartoon}, this integral has a simple geometrical interpretation: The kink travels from time $0$ to time $t$ at momentum $k'-k$, where it interacts with a $\phi$ particle. This changes its momentum to $k'$, with which it moves forward in time through the periodic boundary back to time $0$. To calculate the integrals (\ref{equ:denominator}) and (\ref{equ:numerator}), we use the saddle point approximation.
The saddle point $k_0$ for Eq.~(\ref{equ:numerator})  is found by minimising the action
\begin{equation}
\label{equ:pointaction}
S(k') =   E_{k'} (T-t) + E_{k'-k} t - MT
\end{equation}
for given $t$. By approximating the integral by a Gaussian around the saddle point, we obtain
\begin{multline}
\label{equ:finalcorr}
\left< \phi(0,k)\phi(t,q) \right>
\\=\frac{2\pi\delta(k+q)}{L}  \sqrt{\frac{T}{M}} \frac{v^2|f(k_0,k_0-k)|^2}{\sqrt{E_{k_0-k} E_{k_0} S''(k_0)}} e^{-S(k_0)}.
\end{multline}
This approximation is only valid when the action is sufficiently peaked and
well approximated by a Gaussian.
The latter assertion requires
\begin{equation}
\label{eq:derivativecondition}
S''(k_0)^2 \gg S^{(4)}(k_0).
\end{equation}
This implicitly imposes a lower limit for the system size $T$ in the time direction, so for higher $k$ we need to use larger lattices. As usual, the lattice size also has to be larger than any inverse mass, including the kink mass $M$.

Finally, we note that because $\phi$ is real and  the kink has odd parity, the form factor is odd and purely imaginary. 
Therefore we can use Eq.~(\ref{equ:finalcorr}) to determine it from the field correlator, up to a sign.
For given $k$ and $t$, we obtain the saddle point $k_0$ by minimising Eq.~(\ref{equ:pointaction}), and
the form factor for rapidity difference $\beta$ is given by
\begin{multline}
\label{equ:fresult}
f(\beta)=
f(k_0,k_0-k)\\
=\pm i\frac{\sqrt{\left< \phi(0,k)\phi(t,-k) \right>}}{v} \left(\frac{M E_{k_0-k} E_{k_0} S''(k_0)}{T}\right)^{1/4} e^{S(k_0)}.
\end{multline}
where
\begin{equation}
\beta=\mathrm{arcsinh}\frac{k_0}{2M} - \mathrm{arcsinh}\frac{k_0-k}{2M}.
\end{equation}

While Eq.~(\ref{equ:fresult}) is an approximation, it becomes exact for sufficiently large $T$ as discussed above; one must also satisfy Eq.~(\ref{eq:timesepcondition}) by excluding small $t$.

We tested this result by calculating the form factor near the critical point using lattice Monte Carlo simulations.
The Euclidean lattice action for the theory (\ref{eq:lagrangian}) is given in lattice units by
\begin{multline}
\label{eq:latticeaction}
S = \sum_{\mathbf{x}} \left[ -\sum_{\mu=1}^2 \phi(\mathbf{x})\phi(\mathbf{x}+\mathbf{\hat{\mu}}) \right. \\
 \left. +\left(2 - \frac{m^2}{2}\right)\phi(\mathbf{x})^2 + \frac{\lambda}{4!}\phi(\mathbf{x})^4 \right]
\end{multline}
with $\lambda=0.6$. Square lattice sizes of $L=T\in \{125,\;250,\;375\}$ were used. A kink is created by imposing antiperiodic boundary conditions in the space direction, $\phi(x+L,t)=-\phi(x,t)$.
This also leads to discretisation of momentum,
$k= (2n+1)\pi/L$. 

We measured the momentum space unequal-time field correlator at various time separations and then used Eq.~(\ref{equ:fresult}) to calculate the form factor for various rapidities. At the strong couplings used here, the correlator measurements are reliable even at very long distance thanks to the hybrid Monte Carlo algorithm which was helpful in fighting critical slowing down and thermalizing long-distance modes efficiently.
In principle, Eq.~(\ref{equ:fresult}) gives the form factor for a range of $\beta$ from a single choice of parameters $\{k,L,m^2\}$
because the same simulation gives the correlator for all values of the time separation $t$. However, these 
values are strongly correlated; we report only one data point per combination with a quoted error obtained from a bootstrap resampling of all measurements~\cite{Efron1982}.

Given Eq.~(\ref{eq:timesepcondition}), the time separation $t$ has to be long enough that excited states and two-particle states are suppressed sufficiently. This happens when $t\gtrsim 1/2M$. However, at greater distances statistical noise starts to grow. Therefore, we select the value of $t$ with the smallest statistical error within the permitted range.

In addition to the field correlator, Eq.~(\ref{equ:fresult}) also involves the vacuum expectation value $v$ of $\phi$, and the kink mass $M$.
We measured $v$ using simulations with periodic boundary conditions. We take the ensemble average $v = \left<\left| \sum_{x\in V} \phi(x) \right|\right>$ of the average field's absolute value.

\begin{figure}[tb]
\includegraphics[scale=0.325,angle=270]{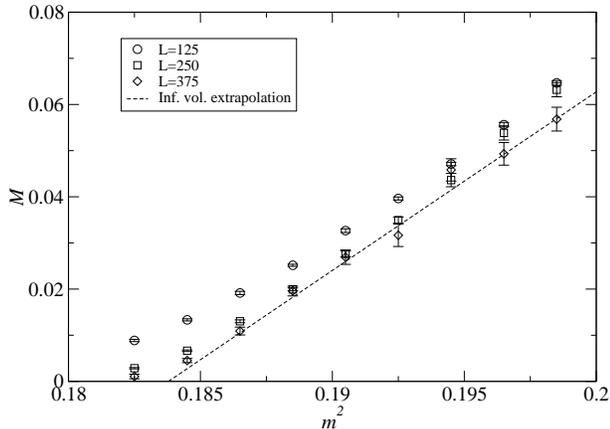}
\caption{\label{fig:masses} The kink mass $M$ as a function of $m^2$ in different volumes. The data agree with the known results for the Ising universality class~\cite{Abraham}: extrapolating to the infinite-volume limit, the dependence on $m^2$ is linear.}
\end{figure}

To obtain the kink mass $M$, we again used Eq.~(\ref{equ:fullres}) and the saddle point approximation, but this time
taking the lowest available momentum, $k=\pi/L$, leaving the higher momentum measurements as independent datasets for study of the form factor. Then, as long as $k\ll M$, Eq.~(\ref{equ:pointaction}) simplifies
and we find the saddle point $k_0=kt/T$ for arbitrary $t$.
We can, therefore, apply the saddle point approximation analytically, and we find~\cite{Rajantie:2009bk}
\begin{equation}
\langle \phi(0,k)\phi(t,q)\rangle \propto e^{-\sqrt{M^2 + k_0^2}t - \sqrt{M^2 + \left(k-k_0\right)^2}(T-t)+MT}.
\end{equation}
We then fit the $k=\pi/L$ correlator data to this expression to obtain the mass $M$ with a bootstrap error. The results of this fitting are shown in Fig.~\ref{fig:masses}. Alternatively, the mass could also be calculated from the free energy difference between the kink and vacuum sectors~\cite{Groeneveld:1980tt}.

This way, we have measured all the quantities that appear in Eq.~(\ref{equ:fresult}). We can calculate the form factor $f(\beta)$ at a wide range of rapidities for different momenta and different values of $m^2$. Finally, we have checked the consistency of the saddle point approximation leading to Eq. (\ref{equ:finalcorr}) when Eq. (\ref{eq:derivativecondition}) is satisfied. 

\begin{figure}[tb]
\includegraphics[scale=0.325,angle=270]{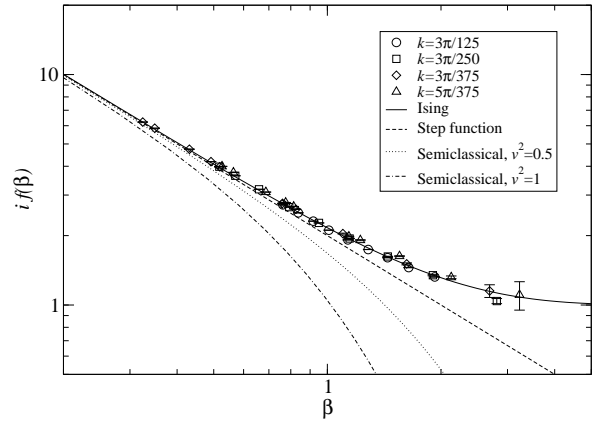}
\caption{\label{fig:formfactor} The form factor as a function of the rapidity difference. Measurements are shown for several lattice sizes at various momenta, and the form factors for the Ising model and for a semiclassical kink are also shown.}
\end{figure}

The results in the critical regime are shown in Fig.~\ref{fig:formfactor}, together with the exact Ising model result (\ref{equ:fIsing}) for comparison. The agreement is very good.
This demonstrates that we can calculate the kink form factor reliably even at strong coupling where
perturbative approaches fail. 
Unlike the Ising model, the scalar field theory (\ref{eq:lagrangian}) in which we carried out the calculation is not exactly solvable, and we made no use of any special features of the theory. Therefore we expect that the same method will work equally well in other theories.

We have shown how to calculate soliton form factors nonperturbatively from field correlation functions measured in lattice field theory simulations.
The approach can be applied directly to other theories with kinks, such as the Sine-Gordon model, 
and generalisation to other theories where twisted boundary conditions can create topological solitons~\cite{Kajantie:1998zn,Davis:2000kv,Edwards:2009bw} should be straightforward. 

In more complicated theories, one will obtain several form factors which describe interactions of the soliton with different particle species.
In the case of point-like solitons, such as 't Hooft-Polyakov monopoles in 3+1-dimensional gauge field theory~\cite{Rajantie:2005hi}, the calculation will follow the same lines. Generalisation to extended solitons, such as domain walls, strings or higher-dimensional membranes is less trivial but should still be possible. For example, this will make it possible to study non-perturbatively the quantum mechanical properties and interactions of cosmic strings. 

The authors would like to thank P.~Dorey and B.~Hoare for useful comments. This work was supported by STFC and made use of the Imperial College HPC Service.

\newpage

\bibliography{kink}
\end{document}